\documentclass[aps,prb,amsmath,twocolumn,amssymb,titlepage,superscriptaddress,showpacs]{revtex4-1}	
\usepackage[english]{babel}
\usepackage{graphicx}
\usepackage{amsmath}
\usepackage{amssymb}
\usepackage{epstopdf}
\usepackage{amsfonts}
\usepackage{bm}
\usepackage{mathdots}
\usepackage{times}
\usepackage{mathtools}
\usepackage{hyperref}

\usepackage{color}

\newcommand{\astcycl}{\mathrlap{\kern0.085em{\circlearrowright}}\ast}
\newcommand{\taucycl}{\mathrlap{\kern0.42em{\bullet}}\circlearrowright}

\begin{document}
\title{Two-channel Kondo physics in a periodically driven single-impurity Anderson model}
\author{Martin Eckstein}
\affiliation{Department of Physics, University of Erlangen-N\"urnberg, 91058 Erlangen, Germany}
\affiliation{Max Planck Institute for the Structure and Dynamics of Matter 22761 Hamburg, Germany}
\author{Philipp Werner}
\affiliation{Department of Physics, University of Fribourg, 1700 Fribourg, Switzerland}


\begin{abstract}
We investigate a quantum dot (Anderson impurity) coupled to metallic leads, with a time-periodic voltage bias across the device. Using a time-dependent Schrieffer-Wolff transformation, we show that the Floquet Hamiltonian of the model can be mapped onto a two-channel Kondo model, in which the impurity is screened by separate conduction bands corresponding to parity-even and odd superpositions of the metallic leads. By changing the frequency and amplitude of the perturbation, one can tune the system to a quantum critical point with symmetric coupling of the impurity to both channels. For the understanding of the driven state, energy absorption from the drive must be considered:  Although the absorption at the impurity is balanced by the energy flow into the conduction band, locally it leads to non-thermal distribution functions which can have a detrimental effect on the Kondo physics. However, a numerical simulation demonstrates that the absorption can be systematically suppressed at a fixed value of the induced couplings by increasing the frequency, so that the time-averaged dynamics in the driven one-channel Anderson model can be used to study the critical behavior of the two-channel model.
\end{abstract}
\maketitle

\section{Introduction}

An intriguing route to control the collective behavior of quantum many-particle systems is to manipulate Hamiltonians and states by time-periodic perturbations. This concept is widely used in experiments with ultra-cold atoms, e.g., to design artificial gauge fields,\cite{Goldman2014} topologically non-trivial band-structures,\cite{Oka2009,Kitagawa2011,Jotzu2015} or frustrated spin models.\cite{Struck2011} Even states with no direct equilibrium analog, such as time-crystals,\cite{Else2016} have been proposed. In solids, periodic driving with strong laser fields can generate Floquet-Bloch bands,\cite{Wang2013} or light-induced many-body interactions such as spin and orbital exchange\cite{Mentink2015,Mikhaylovskiy2015,Eckstein2017,Arenas2017} or superconducting pairing.\cite{Knap2016,Komnik2016,Coulthard2016}

The stroboscopic dynamics under the effect of the driving is described by the so-called Floquet Hamiltonian, and the most interesting situation arises when the latter contains terms which are different from, or even competing with interactions in the un-driven Hamiltonian. In the present paper we will encounter this situation for a quantum dot between metallic leads. In equilibrium and in the strong-coupling regime (Coulomb blockade), the system  is described by a Kondo model, in which a single electron on the impurity with spin $\bm S$  is coupled to the spin density $\bm s_{e}$ of the conduction band via an antiferromagnetic exchange interaction $J_{e} \bm S\cdot\bm s_{e}$. When a time-periodic bias is applied across the impurity, however, spin-exchange processes which involve virtual absorption and re-emission of an  {\em odd} number of photons from the external field act as if the impurity were  coupled to an additional conduction band with spin $\bm s_{o}$, which is distinct from the original channel because of its symmetry. The periodically driven model becomes equivalent to a two-channel Kondo model (2chKM), in which two conduction bands separately screen the impurity, $ \bm S\cdot(J_{e} \bm s_{e}+J_o\bm s_o) $.

The multi-channel Kondo problem\cite{Nozieres1980,Cox1998} gives rise to intriguing physics when the impurity spin becomes over-screened by the conduction electrons. In the  2chKM, the point of equal couplings $J_{e}=J_{o}$ corresponds to a quantum critical point with non Fermi-liquid behavior. The 2chKM can be realized with interacting leads that are made distinct by their charging energy.\cite{Oreg2003,Potok2007,Keller2015}  In alternative proposals, which are based on phonon-mediated exchange interactions\cite{daSilva2009} or a dc bias across the impurity,\cite{Coleman2001} the distinct screening channels correspond to different conduction-band symmetries.  In all cases, however, the controlled manipulation of the couplings remains a challenge. Under the effect of a dc bias the system is in a current carrying steady state where decoherence processes are relevant.\cite{Rosch2001} The proposed setting of this paper provides a symmetry-based realization of the 2chKM in which the exchange couplings can be manipulated by a change of  the frequency and amplitude of the external drive, which may potentially allow to approach quantum critical two-channel Kondo physics using transport setups based on cold atoms.\cite{Krinner2015}

The driven Anderson model has been discussed in different settings, where it was shown that a direct dipole coupling between the localized level and the conduction electrons can induce a Kondo effect.\cite{Iwahori2016a,Iwahori2016b,Nakagawa2015} The study of a driven impurity model touches on a fundamental aspect of driven models in general, i.e,  the energy absorption from the perturbation, which provides a fundamental limitation for the applicability of the Floquet Hamiltonian in bulk systems. The energy absorption can be  either suppressed by going to the limit of high frequencies,\cite{Goldman2014,Bukov2015,Eckardt2015,Mikami2016} where the Floquet Hamiltonian can describe the dynamics over a long period of time (Floquet prethermalization\cite{Canovi2016,Bukov2015PRL}), or it can be balanced with dissipation to additional degrees of freedom.\cite{Babadi2017,Murakami2017} An impurity model is a dissipative quantum system by construction. Although the energy absorption happens only at one site and is thus non-extensive, the relevant {\em  local} distribution functions are strongly modified with respect to the Fermi distribution. This provides a controlled setting to study emerging states in steady state driven models.\cite{Sieberer2016}

The paper is organized as follows: In Sec.~\ref{sec:model} we introduce the driven Anderson model, derive the Floquet Hamiltonian using a time-periodic Schrieffer-Wolff transformation, and present the numerical details for the solution of the one- and corresponding two-channel Anderson model. In Sec.~\ref{sec:results} we numerically solve the dynamics of the Anderson model after a sudden switch-on of the driving, and compare to the numerical solution  of the corresponding two-channel model, both at and away from the critical point. In Sec.~\ref{sec:heating}, we discuss the effect of the local energy absorption, and Sec.~\ref{sec:conclusion} provides a discussion of the results.

\section{Model}
\label{sec:model}
\subsection{Periodically driven Anderson Impurity model}

We start from the Anderson model for a single impurity level at energy $\epsilon_f<0$, coupled to a left (L) and right (R) lead,
\begin{align}
\label{hamiltonian01}
H
&=
H_\text{lead} + H_\text{hyb}+H_\text{dot},
\\
\label{hlead}
H_\text{lead}
&=-t_{0} \sum_{\alpha=L,R} \sum_{i=0}^\infty \sum_{\sigma}
\big( a_{i,\alpha,\sigma}^\dagger a_{i+1,\alpha,\sigma} + h.c.\big),
\\
H_\text{hyb}
&=
v\sum_{\alpha=L,R}  \sum_{\sigma}
\big(c_{\sigma}^\dagger a_{0,\alpha,\sigma} + h.c.\big),
\\
\label{hdot}
H_\text{dot}
&=
U c_{\uparrow}^\dagger c_{\uparrow} c_{\downarrow}^\dagger c_{\downarrow}
+\epsilon_f (c_{\uparrow}^\dagger c_{\uparrow} +c_{\downarrow}^\dagger c_{\downarrow}).
\end{align}
Here $c_{\sigma}^\dagger$ and $a_{i,\alpha,\sigma}^\dagger$ create an electron with spin $\sigma=\uparrow,\downarrow$ on the impurity and on site $i$ of the left ($\alpha=L$) or right ($\alpha=R$) lead, respectively. We represent the lead as a linear chain with nearest neighbor hopping $t_{0}$, and assume a hybridization $H_\text{hyb}$ between the impurity and the first site of each of the two chains. This particular setup is chosen to simplify the notation for the derivation of the Kondo model below; the final numerical results depend only on the hybridization function $\Gamma_{L(R)}(\omega) = 2\pi v^2N_{L(R)}(\omega )$, where $N_{L(R)}(\omega )$ is the density of states at site $0$ of the left (right) lead. The on-site interaction $U$ is taken to be infinite, i.e., only fluctuations between empty and singly occupied states on the impurity are allowed. A finite large $U$ would not alter the general arguments of this work, but the limit $U=\infty$ is numerically less costly. 

To the Hamiltonian \eqref{hamiltonian01} we add a periodic driving term of the form
\begin{align}
H_{dr}=\Delta_0 \sin(\Omega t)  \sum_{i\sigma} (a_{i,L,\sigma}^\dagger a_{i,L,\sigma} - a_{i,R,\sigma}^\dagger a_{i,R,\sigma}),
\end{align}
which describes an oscillating bias $\Delta_0$ across the impurity (but no voltage drop within the leads). It is convenient to remove the driving term by a canonical transformation, which modifies the hybridization to
\begin{align}
H_\text{hyb}
&=
v \sum_{\sigma} 
\big(
e^{i\phi(t)} 
(c_{\sigma}^\dagger a_{0,L,\sigma} 
+
a_{0,R,\sigma}^\dagger   c_{\sigma}
)
+h.c.
\big),
\label{hhyb}
\end{align}
where $\phi(t)=\eta\cos(\Omega t)$ is the Peierls phase with the dimensionless driving amplitude $\eta=\Delta_0/\Omega$.

\subsection{Floquet Hamiltonian}
\label{sec:fws}

In equilibrium, and for small hybridization  $v \ll |\epsilon_f|, |U-\epsilon_f|$, one can use a Schrieffer-Wolff transformation to remove doubly occupied and empty states from the impurity, and thus derive the Kondo model in which the impurity is described by a single spin $\bm S$ with antiferromagnetic exchange coupling  $J\propto v^2/|\epsilon_f|$ to the conduction electron spins. In the periodically driven case, the goal is to achieve an analogous perturbative mapping to a Kondo model for the Floquet Hamiltonian $H_F$, which describes the evolution of the system over one period of the driving. In the following we show that this procedure applied to the driven one-channel Anderson model [Eqs.~\eqref{hamiltonian01}, \eqref{hlead}, \eqref{hdot} and \eqref{hhyb}]  leads to a two-channel Kondo Hamiltonian. 

In analogy to the equilibrium case the idea is to project out empty or doubly occupied states from the {\em time-evolution} of the Anderson impurity model, using a time-dependent unitary transformation $e^{\mathcal{S}(t)}$ to a rotating frame. Such a time-dependent\cite{Sota2016,Eckstein2017} or time-periodic\cite{Bukov2016}  Schrieffer-Wolff transformation has been used in different variants of Hubbard-type lattice models. The Hamiltonian in the rotating frame, $H_\text{rot}(t) = e^{\mathcal{S}(t)} [H(t)-i\partial_t] e^{-\mathcal{S}(t)}$, can be perturbatively constructed in such a way that it does not mix ``high-energy states'' (with an empty or doubly-occupied impurity level)  and ``low-energy states'' (with a singly-occupied impurity level) {\em at any time}. The projection to the low-energy sector then yields a time-dependent (or, for time-periodic driving, a time-periodic) Kondo model. Finally, in the time-periodic case, the driving frequency $\Omega$ is typically large compared to the exchange energy $J$ in the Kondo model (but not necessarily large compared to the high-energy scales $|\epsilon_f|$, $U$), and can be averaged over one period to obtain the  Floquet exchange Hamiltonian. 

The first step to carry out this procedure is a Fourier decomposition $H_\text{hyb}(t)
=
\sum_{n}
e^{-i\Omega n t}
V_{n}
$
of the time-dependent hybridization term Eq.~\eqref{hhyb}. Using the relation ($\tau=2\pi/\Omega$)
\begin{align}
\frac{1}{\tau}
\int_{0}^{\tau} 
dt\,\,
e^{i x \cos(\Omega t+\theta)}
e^{i\omega nt }
=
J_{n}(x)e^{-i\theta n },
\end{align}
for $x>0$, where $J_{n}(x)$ is the Bessel function, with $J_{n}(x)=(-1)^n J_{-n}(x)$, we get
\begin{align}
V_{n}
&=
v
J_{n}(\eta)
\sum_{\sigma} 
\Big[
(c_{\sigma}^\dagger a_{0L\sigma} 
+
a_{0R\sigma}^\dagger   c_{\sigma}
)
+
e^{i\pi n}
\times (h.c.)
\Big]
\nonumber\\
&=
v
J_{n}(\eta)
\sum_{\sigma} 
\Big[
c_{\sigma}^\dagger (a_{0L\sigma} + 
e^{i\pi n}
  a_{0R\sigma} )
+
e^{i\pi n}
\times (h.c.)
\Big].
\label{VFT}
\end{align} 
Note that this relation satisfies $V_{n}^\dagger = V_{-n}$, as required by the hermitian symmetry of $V(t)$. From Eq.~\eqref{VFT} one can see that in even and odd Fourier orders the quantum dot is coupled to even and odd parity combinations of the bath, respectively. 
We therefore introduce the basis change
\begin{align}
\label{b-basis}
b_{i,e/o,\sigma} = \frac{1}{\sqrt{2}}(a_{i,L,\sigma} \pm a_{i,R,\sigma}).
\end{align}
This leaves the Hamiltonian of the leads invariant,
\begin{align}
H_\text{lead}
&=-t_{0} \sum_{\sigma}\sum_{\alpha=e,o}\sum_{i}  
\big(b_{i,\alpha,\sigma}^\dagger b_{i+1,\alpha,\sigma} + h.c.
\big),
\end{align}
and transforms the hybridization to
\begin{align}
\label{hybeven}
V_{n}
&=
\sqrt{2} v
J_{n}(\eta)
\sum_{\sigma} 
(c_{\sigma}^\dagger b_{0,e,\sigma}
+
h.c.
),\,\,(n\text{~even}),
\\
\label{hybodd}
V_{n}
&=
\sqrt{2}v 
J_{n}(\eta)
\sum_{\sigma} 
(c_{\sigma}^\dagger b_{0,o,\sigma}
-
h.c.
 ),\,\,(n\text{~odd}).
\end{align}

This symmetry argument already indicates that after integrating out charge fluctuations by the perturbative transformation in $v$, the impurity spin will be coupled to {\em two} independent baths in the driven case, while in equilibrium there is only one bath. The detailed steps of the perturbative mapping closely follow the time-dependent\cite{Sota2016,Eckstein2017} or time-periodic\cite{Canovi2016,Bukov2016} Schrieffer-Wolff transformations in other models, and are given in the appendix \ref{sec:app01}. The result is the following exchange Hamiltonian,
\begin{align}
H_{F}
=&
-t_{0}
\sum_{i}\sum_{\sigma}\sum_{\alpha=e,o}
\big(
b_{i,\sigma,\alpha}^\dagger
b_{i+1,\sigma,\alpha} + h.c.
\big)
\nonumber
\\
&+
\bm S\cdot \Big[
J_{e}(\eta,\Omega) \bm s_{e,0}
+
J_{o}(\eta,\Omega)\bm s_{o,0}\Big],
\label{2chK}
\end{align}
where  $\bm s_{e(o),0}=
\frac12\sum_{\sigma\sigma'} b_{0,e(o),\sigma}^\dagger {\bm \tau}_{\sigma\sigma'} b_{0,e(o),\sigma'}$ and $\bm S=\frac12\sum_{\sigma\sigma'} c_{\sigma}^\dagger {\bm \tau}_{\sigma\sigma'} c_{\sigma'}$ denote the conduction electron spin at site $0$, the even (odd) bath and the spin at the impurity, respectively (${\bm \tau}$ is the vector of Pauli matrices). The exchange couplings depend on the frequency $\Omega$ and the driving strength $\eta=\Delta_0/\Omega$,
\begin{align}
\label{jbessel}
J_{e(o)}(\eta,\Omega)
=
4v^2
\sum_{l\text{~even(odd)}}
\frac{ J_{|l|}(\eta)^2}{|\epsilon_f| +l\Omega}.
\end{align}
(As mentioned above, the limit $U=\infty$ has been taken in this expression.) 

This result can be intuitively understood: The term proportional to $\frac{ J_{|l|}(\eta)^2}{|\epsilon_f| +l\Omega}$ arises from spin flip scattering processes between the impurity and the bath via a virtual intermediate state which involves an empty impurity (energy $|\epsilon_f|$), and absorption/emission of $l$ photons from the drive (energy $l\Omega$). The summation over all virtual states gives the Floquet exchange coupling, analogous to what has been considered for spin\cite{Mentink2015,Bukov2016} and orbital\cite{Eckstein2017} exchange models. In the present case, absorption or emission of an even or odd number of photons couples the impurity to even or odd parity superpositions of the bath, respectively, which thus appear as separate screening channels for the impurity spin. 

\begin{figure}[tbp]
\centerline{\includegraphics[width=\columnwidth]{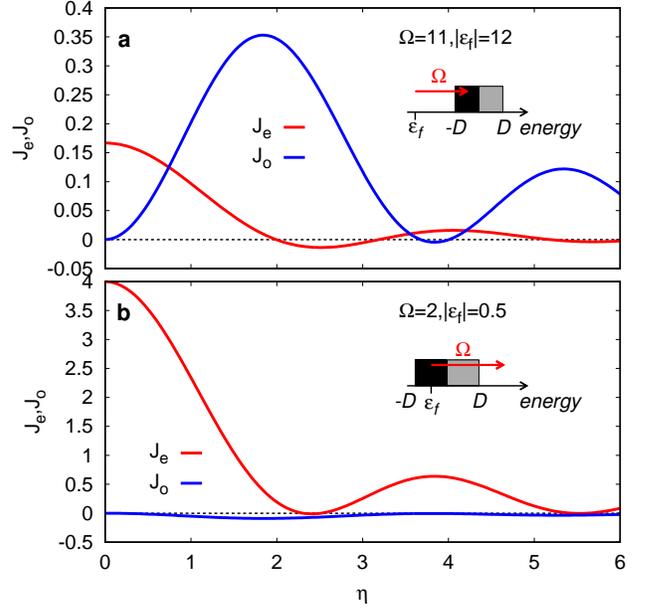}}
\caption{
Exchange interactions $J_e(\eta,\Omega)$ and $J_o(\eta,\Omega)$ in the driven case [c.f.~Eqs.~\eqref{jbessel}]. 
a) Antiferromagnetic coupling $J_o$, for $\Omega<|\epsilon_f|$. 
b) Ferromagnetic coupling $J_o$, for $\Omega>|\epsilon_f|$. 
The insets show schematic energy diagrams for the two cases, where $\Omega$, $\epsilon$, and the bandwidth $D$ are chosen such that there is no linear energy absorption (see discussion in Sec.~\ref{sec:heating}).}
 \label{fig:j}
\end{figure}

Figure \ref{fig:j} shows the exchange couplings as a function of the driving strength $\eta$ for selected parameters $\epsilon_f$ and $\Omega$.  For $\Omega>|\epsilon_f|$,  $J_o$ is ferromagnetic (Fig.~\ref{fig:j}b), which is also understood from Eq.~\eqref{jbessel}: For each $|l|$, the negative term $J_{|l|}(\eta)^2/(|\epsilon_f|-l\Omega)$ is then larger in magnitude than the positive term $J_{|l|}(\eta)^2/(|\epsilon_f|+l\Omega)$. For $\Omega>|\epsilon_f|$, the driven model is therefore a two-channel model where a ferromagnetic coupling competes with an antiferromagnetic coupling (Fig.~\ref{fig:j}b). In contrast, for $\Omega<|\epsilon_f|$, one can have antiferromagnetic coupling in both channels in a large window of driving frequencies (Fig.~\ref{fig:j}a). Furthermore the couplings can be made equal in magnitude, to bring the Hamiltonian into the quantum critical regime of the two-channel Kondo model. In section \ref{sec:results} we will investigate under which conditions this two-channel Kondo model indeed gives a correct description of the dynamics and the steady state of the driven one-channel Anderson model. 

\subsection{Numerical solution: Single impurity Anderson model}
\label{one-channel-siam}

For the numerical studies below we consider a single impurity Anderson model in which the right and left lead correspond to a box density of states $N_{D}(\omega)=\frac{1}{2D}\Theta(D^2-\omega^2)$ with half-bandwidth $D$. This model is defined by the action
\begin{align}
\mathcal{S}
=
-i
\int_C dt H_\text{dot}(t)
-i
\int_C dt dt'
\sum_{\sigma}
c_{\sigma}^\dagger (t)
\Delta(t,t')
c_{\sigma}(t')
\label{onechannelS}
\end{align}
on the $L$-shaped Keldysh contour (for an introduction to the Keldysh-formalism, see, e.g., Ref.~\onlinecite{aoki2014rev}),
where the hybridization function is given by the sum of contributions from the left and right lead, $\Delta(t,t')=\frac{1}{2}\big(\Delta_L(t,t')+\Delta_R(t,t')\big)$, with 
\begin{align}
\label{delta1c}
&\Delta_\alpha(t,t') = V_\alpha(t) \Delta_{D}(t,t') V_\alpha(t')^*, 
\\
&V_L(t)=ve^{i\phi(t)},\,\,\,
V_R(t)=ve^{-i\phi(t)},
\\
\label{kdrive}
&\phi(t) = \eta(t) \cos(\Omega t).
\end{align}
Here $\Delta_{D}(t,t')$ is the Green's function for an equilibrium bath with density of states $-\frac{1}{\pi}\Delta^{ret}_{D}(\omega+i0)=N_{D}(\omega)$, and $\phi(t)$ is the Peierls phase for the driving strength $\eta(t)$, corresponding to a voltage bias $\partial_t \phi(t)$ across the impurity. The model is solved using the time-dependent strong-coupling expansion,\cite{eckstein2010} taking $U=\infty$. In this limit, the second order diagrams (one-crossing approximation) vanish, and we restrict ourselves to the lowest order of the expansion (non-crossing approximation).

\subsection{Numerical solution: Two-channel Anderson model}

We will compare the solution of the driven Anderson model to a two-channel Anderson model, i.e., an extended Anderson model which has the two-channel Kondo model as its strong coupling limit.\cite{Cox1993} This model is given by
\begin{align}
H^{2ch}
&=
\sum_{k\sigma\alpha}
\epsilon_{k\alpha} a_{k,\alpha,\sigma}^\dagger a_{k,\alpha,\sigma} 
+\epsilon_f\sum_{\sigma} f_{\sigma}^\dagger f_{\sigma}
\nonumber\\
&+\sum_{k,\sigma,\alpha}
\big(
V_{k\alpha}   f_{\sigma}^\dagger b_{\bar\alpha} a_{k,\alpha,\sigma} + h.c..\big),
\label{2chsiam}
\end{align}
where $\alpha$ runs over $M=2$ flavors, and $b_{\alpha}$ is an additional bosonic degree of freedom. The model is solved with the infinite-$U$ constraint $\sum_{\alpha} b_\alpha^\dagger b_\alpha+ \sum_{\sigma} f_\sigma^\dagger f_\sigma=1$, i.e., the impurity is either occupied by exactly one fermion or by exactly one boson. In a strong-coupling expansion, virtual intermediate states with zero fermions thus carry the additional flavor index of the auxiliary boson, so that the strong-coupling limit corresponds to an $M$-channel Kondo model.\cite{Cox1993}

After integrating out the bath, the action is given by
\begin{align}
\mathcal{S}_{2ch}
&=
-i
\int_C dt H_\text{dot}^{2ch}(t)
\nonumber\\
&-i
\sum_{\alpha,\sigma}\int_C dt dt'
f_{\sigma}^\dagger b_{\bar\alpha} (t)
\Delta_{\alpha}^{2ch}(t,t')
b_{\bar\alpha}^\dagger f_{\sigma}(t'),
\label{twochannelS}
\end{align}
where the dot-Hamiltonian incorporates the infinite $U$ constraint. The numerical solution of this model follows from a direct application of the strong-coupling expansion on the Keldysh contour \cite{eckstein2010,Cox1993}. Below we will compare the numerical solution of the two-channel model Eq.~\eqref{twochannelS}  with the same box-density of states as in Eq.~\eqref{onechannelS},
\begin{align}
\label{2chsiambath}
\Delta^{2ch}_\alpha(t,t') = v_\alpha^{2ch}(t) \Delta_{D}(t,t') v_\alpha^{2ch}(t')^*,
\end{align}
and suitably chosen couplings (see below) to the dynamics of the driven one-channel model defined by Eq.~\eqref{onechannelS}. 

\section{Results}
\label{sec:results}

\subsection{General setup}

In this section we numerically study the driven one-channel Anderson model (1chAM) with half-bandwidth $D=2$ and a voltage bias $\eta\Omega\sin(\Omega t)$ (Peierls phase $\phi(t)=\eta\cos(\Omega t)$) which is switched on at time $t=0$ [Eqs.~\eqref{onechannelS} to \eqref{kdrive}]. The Floquet Schrieffer-Wolff analysis presented in section \ref{sec:fws} suggests that the dynamics of this model, and the steady state reached under periodic driving, is described by a two-channel Kondo model (2chKM) \eqref{2chK} with exchange interaction \eqref{jbessel}. To confirm this, we compare the dynamics obtained from the 1chAM to a two-channel Anderson model (2chAM) [Eqs.~\eqref{twochannelS} and \eqref{2chsiambath}], with couplings $v_{1,2}^{2ch}$ that are chosen such that the exchange interactions $J_{1,2}^{2ch}$ in the 2chKM derived from the 2chAM match the exchange interactions \eqref{jbessel} in the Floquet Hamiltonian of the 1chAM, i.e., 
\begin{align}
J_1^{2ch}=J_e(\eta,\Omega), \,\,\,
J_2^{2ch}=J_o(\eta,\Omega).
\end{align}
Because all exchange couplings scale with $v^2$, one must satisfy $\big(v_1^{2ch}/v_2^{2ch}\big)^2=J_e(\eta,\Omega)/J_o(\eta,\Omega)$. Furthermore, we know that in the un-driven case the 1chAM [Eqs.~\eqref{onechannelS} to \eqref{kdrive}] with hybridization $v$ is equivalent to the 2chAM [Eqs.~\eqref{twochannelS} and \eqref{2chsiambath}] with $v_1^{2ch}=v$, $v_2^{2ch}=0$. 
To confirm the prediction from the Floquet Schrieffer-Wolff transformation, one thus has to compare the driven 
1chAM (with driving $\eta(t)$ and hybridization $v$) to the 2chAM with hybridizations
\begin{align}
\label{2chparams}
v_1^{2ch}(t)
&=v\sqrt{J_e(\eta(t),\Omega)/J_e(0,\Omega)},
\\
v_2^{2ch}(t)
&=v\sqrt{J_o(\eta(t),\Omega)/J_e(0,\Omega)}.
\label{2chparams2}
\end{align}
To make the actual comparison, we focus on the impurity contribution to the magnetic susceptibility $\chi=\partial \langle S_z \rangle //\partial B_z$, which provides a good characterization of the dynamics of the model in the Kondo regime and in the quantum critical regime of the two-channel model.\cite{Andrei1984,Tsvelick1984,Sacramento1991}

As mentioned in the introduction, the parameters $\epsilon_f$, $D$, and $\Omega$ must be chosen such that energy absorption by generation of particle-hole pairs is low. To avoid direct excitations from the level at $\epsilon_f$ to the unoccupied density of states of the bath, $\epsilon_f+\Omega$ must lie either (i), in the occupied part of the density of states, $\epsilon_f + \Omega<0$ (inset in Fig~\ref{fig:j}b), or (ii), above the conduction band, $\epsilon_f+\Omega >D$ (inset in Fig~\ref{fig:j}a). We will focus on the setup (i), in which both couplings $J_e$ and $J_o$  are antiferromagnetic, and the system can be brought to the critical region $J_e\approx J_o$. In the remainder of Sec.~\ref{sec:results} we choose $\epsilon_f=-12$, $D=2$, $\Omega=11$, and $v=3$. The dependence on $\epsilon_f$ will be investigated in Sec.~\ref{sec:heating}.

\subsection{Two-channel Anderson model: equilibrium}

\begin{figure}[tbp]
  \centerline{
  \includegraphics[width=\columnwidth]{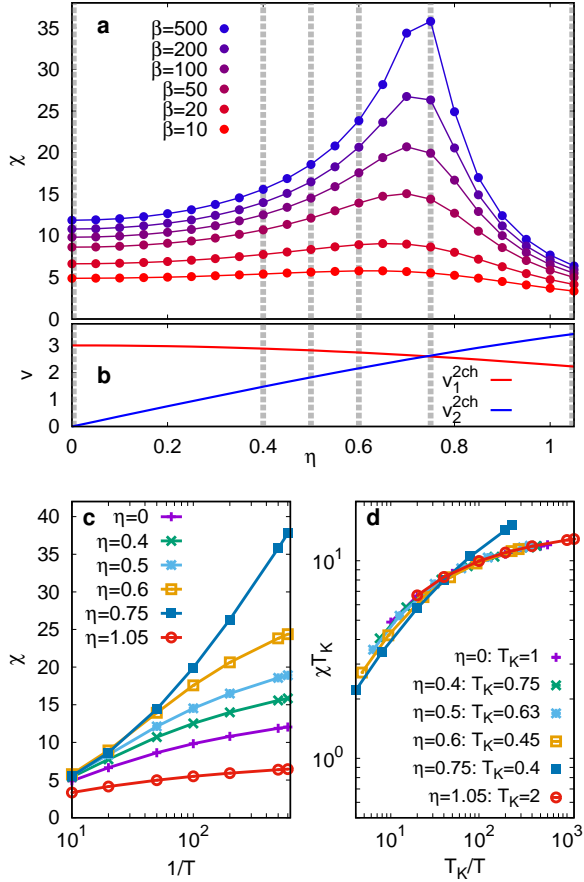}%
    }
  \caption{
a) Spin-susceptibility $\chi$ of the 2chAM model with $\epsilon_f=-12$, $D=2$, along a path through parameter space $v_{1,2}^{2ch}$, for different inverse temperatures $\beta$. The path is parametrized by taking $0\le \eta \le 1.1$ and $\Omega=11$ in Eqs.~\eqref{2chparams} and \eqref{2chparams2}. The corresponding parameters $v_{1,2}^{2ch}$ are shown in panel b. c)  Same data as in a), plotted as a function of $\beta$ for selected values of $\eta$ (see vertical dotted lines in panels a and b). d) Data of panel c) on a double-logarithmic plot. The axes are rescaled with a different factor $T_K(\eta)$ for each curve, as indicated in the legend.
}
\label{fig:equi}
\end{figure}

Before presenting the time-dependent data, let us analyze the equilibrium solution of the 2chAM [Eqs.~\eqref{twochannelS}-\eqref{2chsiambath}] around the critical point $v_1^{2ch}=v_2^{2ch}$, to confirm that with the choice of parameters $\epsilon_f=-12$, $D=2$, $v=3$, $\Omega=11$ in the driven 1chAM one should be able to access the specific dynamics of the 2chKM by varying the driving amplitude $\eta$. Figure~\ref{fig:equi}a shows the equilibrium susceptibility $\chi$ of the 2chAM for different temperatures along a path through parameter space $v_{1,2}^{2ch}$. The values  $v_{1,2}^{2ch}$ along the path, which are displayed in the the lower panel of Fig.~\ref{fig:equi}, correspond to a variation of the parameter $\eta$ in Eqs.~\eqref{2chparams} and \eqref{2chparams2}. $\chi$ is computed explicitly by solving the 2chAM with a magnetic field $B_z$ and taking the numerical derivative $\partial \langle S_z\rangle/\partial B_z$ at $B_z=0$.  The susceptibility $\chi$ shows a strong increase close to the point $v_1^{2ch}=v_2^{2ch}$  for low temperatures, which is the first indication of the critical behavior. In Fig.~\ref{fig:equi}c, we plot $\chi$ as a function of temperature for selected points of the path (see vertical dotted lines in Fig.~\ref{fig:equi}a). In the Kondo model, one expects a free-spin behavior $\chi\sim 1/T$ for large temperatures, and a saturation $\chi\sim 1/T_K$ below the Kondo temperature. In contrast, at the critical point the susceptibility should divergence like  $\chi\sim \log(1/T)$ at low temperatures.\cite{Andrei1984,Tsvelick1984,Sacramento1991}  The data of Fig.~\ref{fig:equi}c are consistent with this phenomenology. For parameters away from the critical point, $\chi(T)$ tends to saturate, while at the critical point ($\eta=0.75$) we observe a logarithmic increase $\chi(T)\propto \log(T)$.

If the system is in the Kondo regime, the susceptibility $\chi(T)$ should fall on a universal line when $T_K\chi$ is plotted as a function of $T/T_K$. This scaling indeed works reasonably well in the parameter regime covered by our investigation for $v_{1}^{2ch}\neq v_{2}^{2ch}$, and fails at the critical point $v_{1}^{2ch}= v_{2}^{2ch}$ (see Fig.~\ref{fig:equi}d). Note that we do not have an analytical expression for the Kondo temperature $T_K(v_1^{2ch},v_1^{2ch})$ for the 2chKM, but treat the parameter $T_K(\eta)$ in  Fig.~\ref{fig:equi}d as  a fit parameter to obtain the best scaling plot. This implies that $T_K$ is only determined relative to the value in the one-channel model $(\eta=0)$. Nevertheless, the plot confirms the decrease of $T_K$ towards the critical point. (To confirm the precise dependence $T_K \sim (J_1^{2ch}-J_2^{2ch})^2$ we would have to study much lower temperatures, which is not in the scope of this work, where we focus on the driven state.)

\begin{figure}[tbp]
\centerline{\includegraphics[width=\columnwidth]{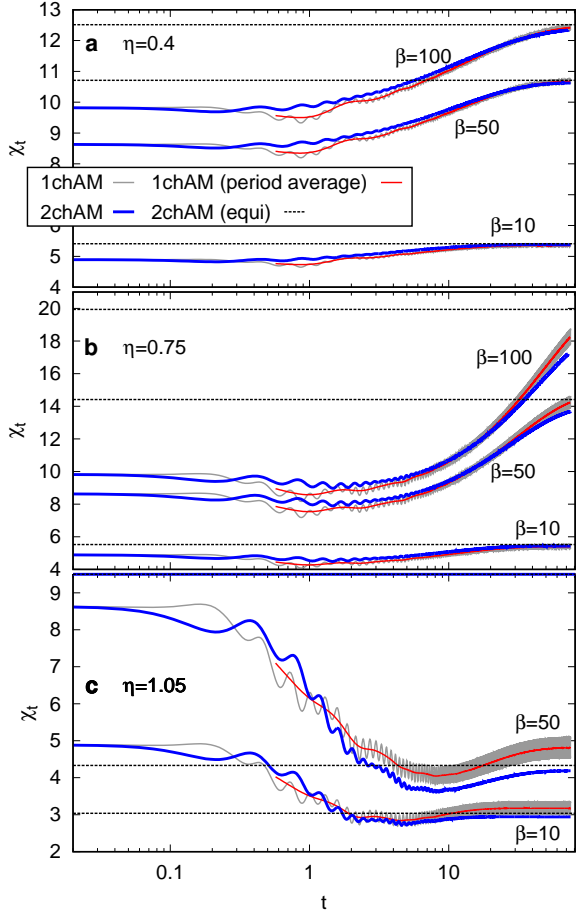}}
  \caption{
Susceptibility $\chi_t(t)$ computed from the time-dependent and initial state expectation value of the spin in the presence of a small field $B_z=10^{-4}$ [Eq.~\eqref{chi}]. In all panels, the thin grey line shows the dynamics in the 1chAM for $\epsilon_f=-12$, $D=2$, $v=3$, and bath temperatures as indicated by the labels, where the ac bias with $\Omega=11$ and amplitude $\eta=0.4$ (a), $\eta=0.75$ (b), and $\eta=1.05$ (c) is switched on at time $t=0$. The thin red curves show the same data averaged over one period, and the bold blue line shows $\chi_t(t)$ in the 2chAM with corresponding parameters, i.e., for a quench from $v_{12}^{2ch}(\eta=0,\Omega)$ to $v_{12}^{2ch}(\eta,\Omega)$ [Eqs.~\eqref{2chparams} and \eqref{2chparams2}]. The horizontal dotted lines correspond to the equilibrium value of $\chi$ in the 2chAM at the final coupling $v_{12}^{2ch}(\eta,\Omega)$.
}
\label{fig:quench1}
\end{figure}

\subsection{Driven one-channel Anderson model}

The analysis of the previous section confirms that with the choice of parameters $\epsilon_f=-12$, $D=2$, $\Omega=11$, $v=3$ we should be able to access the dynamics of a 2chKM by varying the driving amplitude $\eta$ in the 1chAM. The simplest protocol is to suddenly switch on the ac bias, i.e., $\eta(t)\equiv\eta$ for $t\ge 0$ in Eq.~\eqref{kdrive}, and compare the dynamics to the 2chAM where the couplings are quenched from $v_{1}^{2ch}=v$, $v_{2}^{2ch}=0$ for $t<0$ to $v_{1}^{2ch}(\eta,\Omega)$, $v_{2}^{2ch}(\eta,\Omega)$ for $t>0$  [c.f.~Eqs.~\eqref{2chparams} and \eqref{2chparams2}]. During the time evolution we apply a small magnetic field $B_z=10^{-4}$ and analyze the time-evolution of the spin $\langle S_z\rangle$ on the impurity,
\begin{align}
\label{chi}
\chi_t(t)\equiv\langle S_z(t) \rangle / B_z.
\end{align}
(We confirmed that in the parameter range of this study, $\langle S_z(t) \rangle $ is linear in $B_z$, so that $\chi_t$ for $B_z=10^{-3}$  and $B_z=10^{-4}$ would be indistinguishable on the scale of the plots.) If the system would behave adiabatically (or is in equilibrium), $\chi_t$ would  be given by the equilibrium magnetic susceptibility $\chi$. Also away from a quasi-equilibrium state this quantity can be used to directly compare the dynamics in the driven 1chAM and the corresponding 2chAM. 

The data are shown in Fig.~\ref{fig:quench1}.  For both quenches into the Kondo regime with $J_e>J_o$ (Fig.~\ref{fig:quench1}a) and $J_e<J_o$ (Fig.~\ref{fig:quench1}c), and to the critical point $J_e=J_o$ (Fig.~\ref{fig:quench1}b), we find that the time-evolution of the spin in the driven model closely follows the evolution in the respective 2chAM. Small quantitative differences are expected, since the analytical relations \eqref{2chparams} and \eqref{2chparams2} between the two models have been derived for the strong-coupling limit $v\ll |\epsilon_f|$, while the simulation is done at finite $v$. For the quenches within the Kondo regime, $\chi_t(t)$ first increases and then saturates at long times to a value which is close to the equilibrium value of $\chi$ in the 2chAM at the final parameters $v_{1,2}^{2ch}(\eta,\Omega)$ (dotted horizontal lines). For quenches to the critical point, we cannot reach saturation at low temperatures, and $\chi_t(t)$ keeps increasing logarithmically ($\propto \log (t)$) at long times. The increase of the numerical cost with simulated time prevents us to follow this logarithmical increase over more than an order of magnitude (and thus to rigorously distinguish it from a Kondo effect with a very small Kondo temperature), but nevertheless, within the simulated time interval the 2chAM and the driven 1chAM follow the same qualitative behavior.

\section{Energy absorption from the perturbation}
\label{sec:heating}

As mentioned above, energy absorption from the periodic perturbation can be expected to have a detrimental effect on the Kondo physics, comparable to increasing the bath temperature. In the driven impurity model, energy absorption is eventually balanced by an energy flow into the bath. Since the bath is infinite, the  global energy per site remains unchanged, but nevertheless the {\em local} distribution functions can  be strongly modified with respect to a Fermi distribution. In the following we compute the energy absorption and the local distribution functions as a function of the model parameters, and investigate to what extent the effect of the energy absorption on the local environment of the impurity is similar to an increase of the bath temperature.

\begin{figure}[tbp]
 \centerline{ \includegraphics[width=\columnwidth]{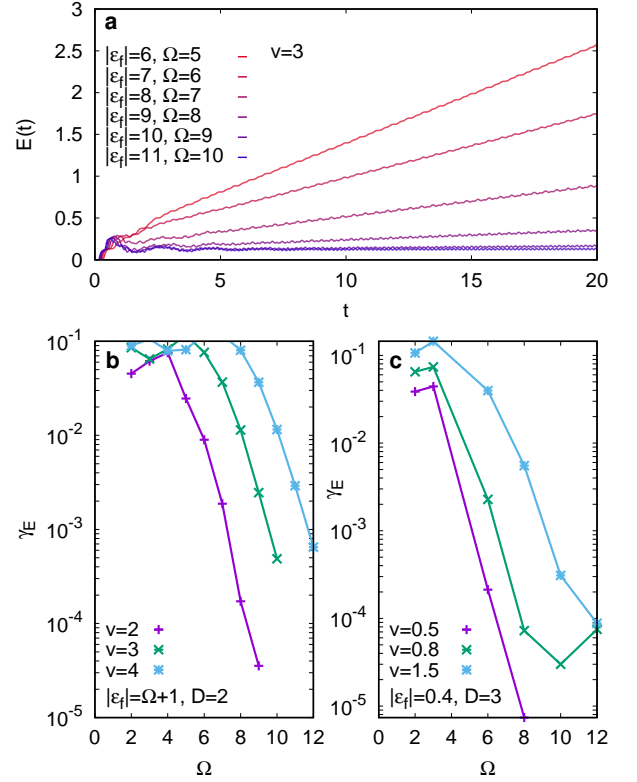}}
 \caption{
a) Absorbed energy for the driven 1chAM ($D=2,v=3,\eta=1,\beta=20$), where we fix $\epsilon_f+\Omega=-1$, while varying $\Omega$. b) Absorption rate $\gamma_E$, obtained from a linear fit $E(t)=\text{const.} + \gamma_E t$ to $E(t)$ in a time window $ 10\le t\le 20\tau $ covering $20$ periods $\tau$ of the perturbation. For low absorption rates ($\sim 10^{-4}$) the fit becomes less accurate because of the oscillating component in $E(t)$. c) Similar analysis as in panels a and b, for a 1chAM with $D=3,\beta=20$, where $\epsilon_f=-0.4$ is kept fixed while $\Omega$ is varied so that $\epsilon_f+\Omega >D$.
}
  \label{fig:absorption}
\end{figure}

\begin{figure}[tbp]
  \centerline{
  \includegraphics[width=\columnwidth]{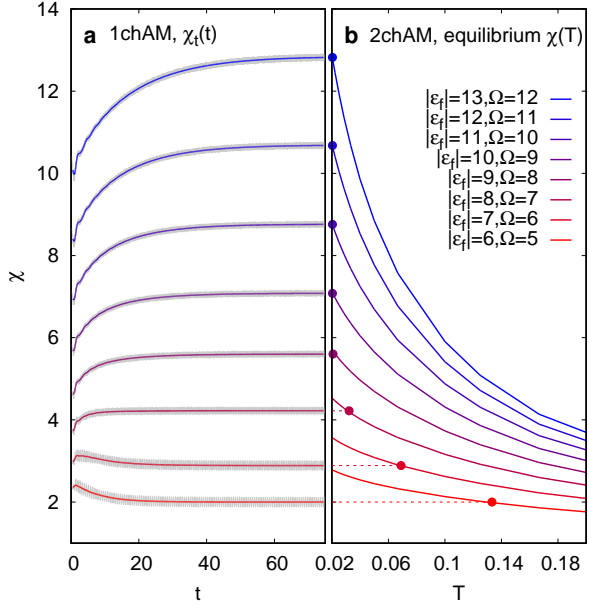}
    }
  \caption{
a)
Susceptibility $\chi_t(t)$ of the 1chAM for $D=2,v=3,\beta=50,\eta=0.4$ and different driving frequencies $\Omega$, where $\epsilon_f+\Omega=-1$ is kept fixed (legend in panel b). Thin grey lines and bold colored lines show the full time-dependence and the period-averaged data, respectively. b) Equilibrium susceptibility $\chi(T)$ of the corresponding 2chAM with  $D=2,v=3$ and couplings $v_{1,2}^{2ch}(\eta=0.4,\Omega)$. Filled dots indicate the effective temperature $T_{\gamma_E}$ for which $\chi(T)$ would equal the susceptibility $\chi_t$ in the driven steady state.
}
  \label{fig:absorption02}
\end{figure}

The absorption rate will depend strongly on $\Omega$, $|\epsilon_f|$, and the bandwidth $D$ of the bath. Linear absorption from an occupied level at energy $-|\epsilon_f|$ to the unoccupied bath density of states at $0\le \epsilon\le D$ is possible for 
\begin{align}
0 \le -|\epsilon_f| +\Omega \le D.
\end{align}
For other frequencies, absorption occurs only via strongly suppressed higher order processes, i.e., either more than one electron-hole pair is generated per absorbed photon, or multiple photons are absorbed per particle-hole pair. To analyze the energy absorption in the 1chAM, we compute the currents $j_R$ and $j_L$ to the right and left lead, which are given by 
\begin{align}
j_{L}(t) =-j_{R}(t)= \text{Re} \big(
[\Delta_{L}\ast G]^<(t,t)/V_{L}(t)\big),
\end{align}
where $G(t,t')=-i\langle T_{C} c^\dagger (t) c(t')\rangle$ is the impurity Green's function and $\ast$ denotes the convolution. With the voltage bias $\eta\Omega \sin(\Omega t)$, the absorbed energy is proportional to 
\begin{align}
E(t) =\eta\Omega \int_0^t ds\, j_L(s)  \sin(\Omega s).
\end{align}
Figure \ref{fig:absorption}a shows the absorption in the setup where $D=2$ sets the energy scale, and $\epsilon_f+\Omega=-1$ is kept fixed within the occupied part of the conduction band, while $\Omega$  (and $\epsilon_f$) is varied (compare inset in Fig.~\ref{fig:j}a). For small frequencies, $E(t)$ increases linearly in time (with an oscillatory component), while in the far off-resonant regime the absorption becomes small. The (period-averaged) energy absorption rate $\gamma_E$ decreases at least exponentially with frequency (Fig.~\ref{fig:absorption}b). A similar $\Omega$-dependence of the absorption rate is shown in Fig.~\ref{fig:absorption}c for the setup in which the energy $\epsilon_f$ of the level is fixed, and $\Omega+\epsilon_f$ lies above the conduction band, compare inset in Fig.~\ref{fig:j}b. Such an exponential behavior  
of the absorption rate is reminiscent of the 
problem of the decay of a high-energy excitation into multi-particle excitations.\cite{Strohmaier2010} In the latter case, a timescale $\gamma\sim \exp(-U/W\log(U/W))$ was found for the decay of a doublon with energy $U$ in the Hubbard model into particle-hole excitations of typical energy $W\ll U$. In the present numerical data, $\gamma_E\sim\exp(-\Omega/D)$ and  $\gamma_E \sim\exp(-\Omega/D\log(\Omega/D))$ would be almost indistinguishable, taking into account that the relative accuracy of rates below $\approx 10^{-4}$ becomes low, because fitting $E(t)$ with a linear increase is then more strongly influenced by the oscillating component of $E(t)$.

In Fig.~\ref{fig:absorption02} we investigate the effect of energy absorption on the dynamics at the impurity. In the left panel, we plot $\chi_t(t)$ for the same setup as Fig.~\ref{fig:absorption}b, and $v=3$, analogous to Fig.~\ref{fig:quench1}a. In the right panel, we compare this to the temperature-dependent equilibrium susceptibility in the 2chAM with corresponding parameters $v_{12}^{2ch}(\eta,\Omega)$ [cf.~Eq.~\eqref{2chparams}]. For large frequencies, i.e., small energy absorption, $\chi_t(t)$ approaches a driven steady-state value corresponding to the equilibrium $\chi$ at the temperature of the bath ($1/\beta=0.02$). For small frequencies ($\Omega\lesssim 8$), the final value of $\chi_t(t)$ in the driven steady state is decreased with respect to the equilibrium value.  As far as the susceptibility is concerned, the energy absorption in the steady state at the impurity site therefore has a similar effect as increasing the temperature to a larger value $T_{\gamma_E}$ (c.f. filled circles in the right panel of  Fig.~\ref{fig:absorption02}).

\begin{figure}[tbp]
  \centerline{
  \includegraphics[width=\columnwidth]{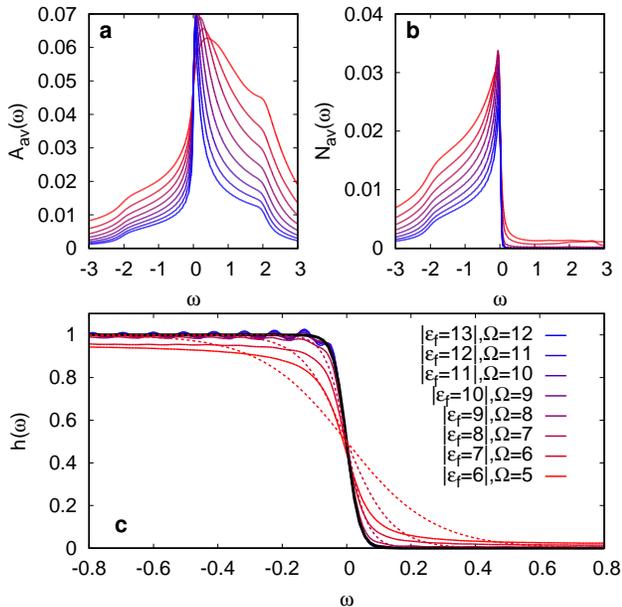}
    }
  \caption{
a) Spectral function Eq.~\eqref{aw} in the driven steady state of the 1chAM, for the same parameters as in Fig.~\ref{fig:absorption02}a (legend in panel c), evaluated at the largest possible simulation time $t=75$. b) Occupied density of states [Eq.~\eqref{nw}] for the same parameters. c) Distribution function $h(\omega)=N_{av}(\omega)/A_{av}(\omega)$ for the same parameters. The solid black line shows the Fermi distribution at $T=1/50$. Dotted lines show the Fermi distribution at $T_{\gamma_E}=1/7.5, 1/14.5, 1/31$ (corresponding to the filled circles in Fig.~\ref{fig:absorption02}b).
}
  \label{fig:absorption03}
\end{figure}

It is therefore instructive to check whether also the distribution functions of the local impurity correlation functions resemble an effective thermal state. In Fig.~\ref{fig:absorption03} we plot, for the same parameters as in Fig.~\ref{fig:absorption02}, the spectral function 
\begin{align}
\label{aw}
A_{av}(\omega,t)= -\frac{1}{\pi }\,\text{Im} \,\,\frac{1}{\tau}\int_{t-\tau}^{t}\!\!\!\!\!\!d\bar t  \int_0^{s_{c}} \!ds\, G^{ret}(\bar t,\bar t-s)e^{i(\omega+i0)s},
\end{align}
and the distribution function
\begin{align}
\label{nw}
N_{av}(\omega,t)= 
\frac{1}{\pi }\,\text{Im}
\frac{1}{\tau} \int_{t-\tau}^{t}\!\!\!\!d\bar t\, \int_0^{s_{c}}\!ds\, G^{<}(\bar t,\bar t-s)e^{i\omega s},
\end{align}
averaged over one period $\tau=2\pi/\Omega$ ($s_{c}=t-\tau$ is a large cutoff in time, due to the finite simulated time window). The spectra show the Kondo resonance around $\omega=0$ (Fig.~\ref{fig:absorption03}a). In addition, we compute the distribution function $h(\omega,t)=N_{av}(\omega,t)/A_{av}(\omega,t)$, shown in Fig.~\ref{fig:absorption03}c. In particular in the regime of large absorption, we observe that $h(\omega,t)$ quickly approaches a time-independent form, so that the curves shown in Fig.~\ref{fig:absorption03}c correspond to the distribution in the driven steady-state. 

For large $\Omega$, the distribution function is given by the Fermi function at the temperature of the bath. (The oscillations are due to the cutoff of the Fourier integrals in Eqs.~\eqref{aw} and \eqref{nw} at the largest simulation time.) For small frequencies and large absorption, however, the distribution functions become highly non-thermal, as indicated by the prominent tails with $h(\omega)<1$ for $\omega<0$ and $h(\omega)>0$ for $\omega>0$. In particular, the distribution functions do neither equal the Fermi function at the bath temperature $1/\beta$ (solid black line), nor at the increased temperature $T_{\gamma_E}$ which would explain the value of $\chi$ in the steady state (taken from Fig.~\ref{fig:absorption02}b). This shows that although the steady state  energy absorption has similar effects on the Kondo effect as an increased temperature, an effective thermal description of the driven state is not possible. 

\section{Conclusion and Discussion}
\label{sec:conclusion}

In conclusion, we have used a time-dependent Schrieffer-Wolff transformation to show that a quantum impurity model driven by an ac bias can display two-channel Kondo physics. By varying the amplitude of the drive, the system can be tuned to the critical point of the two-channel model. We have numerically demonstrated that in a regime where energy absorption at the impurity is low, the time-evolution of the impurity spin $S_z$ in the driven impurity model follows the dynamics of a two-channel model. At the critical point, the spin susceptibility increases logarithmically with time. 

To observe the critical behavior, one must go to a regime in which energy absorption at the impurity is small. Otherwise, the system evolves into a steady state in which the local distribution functions are highly nonthermal and the spin susceptibility is reduced, although the global energy per particle in the bath is unchanged. Energy absorption can be reduced by off-resonant driving, i.e., large $\Omega$. For a potential experimental realization of the critical behavior, it is however not only important to decrease the absorption, but this must be done in such a way that the exchange energies $J$  are not simultaneously decreased in the same way, because energy absorption has a similar decohering effect as an increase of the bath temperature, and the latter must be smaller than $J$ (or even the Kondo temperature $T_K(J)$) to access the critical regime. Figure \ref{fig:absorption}b shows that this is indeed possible: Fixing $\epsilon_f+\Omega<0$, $\gamma_E$ is strongly reduced by increasing $\Omega$. At fixed $v$, the dominant contribution $v^2/|\epsilon_f|$ to the exchange $J_e$ would be also reduced, but by comparing $\gamma_E$ for different couplings $v$ one can see that $\gamma_E$ can be systematically reduced while $v^2/|\epsilon_f|$ is kept constant (compare, e.g., $v=2$, $|\epsilon_f|=4$, $\Omega=3$ ($\gamma_E\approx 5 \cdot10^{-2}$) to $v=3$, $|\epsilon_f|=9$, $\Omega=8$ ($\gamma_E\approx5\cdot10^{-3}$), and the extrapolation of the curve $v=4$ to $|\epsilon_f|=16$, $\Omega=15$ ($\gamma_E\approx6\cdot10^{-5}$)).

A corresponding design of the conduction electron bath could be possible in particular in cold-atom realizations of a quantum transport setting.\cite{Krinner2015} Although this is definitely challenging from an experimental point of view, it would provide an intriguing route to observe quantum critical behavior in such experiments. Even if the temperature of the bath is not yet far below the Kondo-temperature, the critical behavior would manifest itself by a peak in the impurity spin susceptibility as a function of the driving amplitude in the critical regime. One could compare these data to the results of a setting in which the impurity level $\epsilon_f$ is periodically modified with respect to the bath. In contrast to the setup of this paper, the bias between the impurity and the two leads is then  in phase and not out of phase, so that the driven model reduces to a one-channel model with renormalized parameters, which should show no peak of the susceptibility at the same driving amplitude. In theory, even designing different multi-channel models, by coupling more than two baths which are driven by an ac-bias with a relative phase difference could be possible.

\acknowledgements

ME acknowledges support by the DFG within the Sonderforschungsbereich 925 (project B4). PW acknowledges support from FP7 ERC Starting Grant No.~278023. ME and PW thank the Kavli Institute for Theoretical Physics (National Science Foundation Grant No. NSF PHY-1125915) for its hospitality. 

\appendix

\section{Time-dependent Schrieffer Wolff transformation}
\label{sec:app01}

\newcommand{\St}{\mathcal{S}}

To derive an effective Kondo Hamiltonian, one must project out from the dynamics all terms in which the impurity is either empty of doubly occupied, i.e, which do not satisfy the constraint  $n=\sum_{\sigma}c_{\sigma}^\dagger c_{\sigma} = 1$. We will denote the projector on this ``low-energy'' sector of the Hilbert space by $\mathcal{P}_0$, and define $\mathcal{P}_1=1-\mathcal{P}_0$. To do the projection, we look for a {\em time-periodic} unitary transformation $W(t)$ into a rotating frame, in which states with  $n = 1$ are not mixed in time with $n \neq 1$, i.e., the Hamiltonian $H_\text{rot}$ in the rotating frame commutes with $\mathcal{P}_0$. For any time-dependent unitary transformation  $W(t)=e^{\St(t)}$ (parametrized by the antihermitian matrix $\St$), which transforms the wave function like $|\Psi_\text{rot}(t)\rangle = e^{\St(t)} |\Psi(t)\rangle$, the Hamiltonian in the rotating frame is given by
\begin{align}
H_\text{rot}(t) = e^{\St(t)} [H -i\partial_t ]e^{-\St(t)}.
\end{align} 
In the following we write $H=\alpha V+H_0$, where $H_0=H_\text{dot}+H_\text{lead}$ and $V=H_\text{hyb}$ are the Hamiltonian of the decoupled impurity-bath system and the hybridization, respectively, and perform a perturbation expansion in the formal small parameter $\alpha$. A Taylor ansatz $\St=\alpha \St_1 + \alpha^2 \St_2 + \cdots$ yields the series 
\begin{align}
\nonumber
&H_\text{rot}(t)
=
 H_0 
+
\alpha 
\big\{
 V
+
[\St_1, H_0]
+
i\dot \St_1
\big\}
+
\alpha ^2
\big\{
[\St_2, H_0]
\\
&
+
i \dot \St_2
+
[\St_1, V]
+
\frac{1}{2}
[\St_1,i\dot \St_1 + [\St_1, H_0]]
\big\}
+
\mathcal{O}(\alpha^3).
\label{hrotexpansion}
\end{align}
Furthermore, we can make a Fourier decomposition for all time-periodic operators,
\begin{align}
A(t)&=\sum_{n} A^{(n)}e^{-i\Omega nt}, 
\\
A^{(n)}&=
\frac{1}{\tau}
\int_{0}^\tau dt 
A(t) e^{i\Omega n t}.
\end{align}
To first order, the Fourier series of Eq.~\eqref{hrotexpansion} reads
\begin{align}
\label{hrot-FT}
H_\text{rot}^{(n)}
=
\delta_{n,0}H_0
+
\alpha 
\big(
V^{(n)}
+
[\St_1^{(n)},H_0]
+
n\Omega \St_1^{(n)}
\big)
+
\mathcal{O}(\alpha^2),
\end{align}
where we have used that only the hybridization term is time-dependent, i.e., $H_0^{(n)}=\delta_{n,0}H_0 $, and $V^{(n)}$ is given by Eqs.~\eqref{hybeven} and \eqref{hybodd}. To remove the hopping term between the high and low energy sector to lowest order one must satisfy (for each Fourier component)
\begin{align}
\label{order1}
\mathcal{P}_{1(0)}\Big\{
V^{(n)}
+
[\St_1^{(n)},H_0]
+
\Omega n \St^{(n)}
\Big\}
\mathcal{P}_{0(1)}
=0.
\end{align}
We denote the matrix elements of $\St$ between different sectors of the Hilbert space by
$\St_{+} = \mathcal{P}_{1} \St_{1} \mathcal{P}_{0}$
and
$\St_{-} = \mathcal{P}_{0} \St_{1} \mathcal{P}_{1}$ (analogous notation for $V$),
and write Eq.~\eqref{order1} in a basis $E_n|n\rangle=H_0|n\rangle$ of the uncoupled Hamiltonian,
\begin{align}
&\langle m|V^{(n)}_\pm|m'\rangle
+
\langle m|\St_\pm^{(n)}|m'\rangle E_{m'}
\nonumber\\
&
\hspace{5mm}
-
\langle m|\St_\pm^{(n)}|m'\rangle E_{m}
+
\Omega n \langle m|\St_\pm^{(n)}|m'\rangle
=0.
\end{align}
Hence Eq.~\eqref{order1} is satisfied with $\St_1=\St_{+}+\St_{-}$, and 
\begin{align}
\label{S1}
\langle m|\St_\pm^{(n)}|m'\rangle
=
\frac{\langle m|V^{(n)}_\pm|m'\rangle}{E_{m}-E_{m'}-n\Omega}.
\end{align}
It is already clear from this expression that the driving frequency should not allow resonant transitions between impurity and bath sites. Note that because $V$ has no matrix within the high and low energy sector of the Hilbert space, and $H_0$ has no matrix elements between the sectors, the choice \eqref{S1} implies that all first order terms in \eqref{hrot-FT} vanish,
\begin{align}
\label{order1-unprojected}
V(t)
+
[\St_1(t),H_0]
+
i\dot \St_1(t)
=0,
\end{align}
not only the matrix elements \eqref{order1}. 

As a next step, one has to proceed to the second order to derive the effective Kondo Hamiltonian. Using \eqref{order1-unprojected} in the expansion \eqref{hrotexpansion}, we obtain, for the second order terms,
\begin{align}
\alpha^2\Big\{
[\St_2,H_0]+i\dot \St_2 + \frac{1}{2}[\St_1,V]
\Big\}. 
\end{align}
Proceeding as before, all second order terms which mix sector $0$ and $1$ of the Hilbert space are removed by a proper choice of $S_2$. The terms which remain in sector $0$ derive from the last commutator, $\mathcal{P}_0\tfrac12  [\St_1,V]\mathcal{P}_0 $, which therefore gives the exchange Hamiltonian for the impurity spin. Inserting Fourier components, the Hamiltonian in the $00$ sector is 
\begin{align}
H^{(n)}_\text{rot}
&=
\frac{\alpha^2 }{2}
\sum_{l}
\mathcal{P}_0 
(\St_1^{(n-l)}V^{(l)}
-
V^{(l)} \St_1^{(n-l)} 
)
\mathcal{P}_0
\\
&=
\frac{\alpha^2 }{2}
\sum_{l}
(\St_+^{(n-l)}V_-^{(l)}
-
V_+^{(l)} \St_-^{(n-l)}),
\end{align}
where in the second line we have assumed $U=\infty$ for simplicity, such that only fluctuations to the empty state are possible.

Finally, if the driving frequency $\Omega$ is large compared to the relevant energy scales in the low-energy sector (i.e., the exchange interaction), we can perform a high-frequency expansion and compute the time-averaged exchange Hamiltonian,
\begin{align}
H_\text{eff}\equiv
H^{(0)}_\text{rot}
=&
\frac{1}{2}
\sum_{l}
\big(\St^{(-l)}_-V^{(l)}_+-  V^{(l)}_- \St^{(-l)}_+
\big).
\label{almostKondo}
\end{align}
To show that Eq.~\eqref{almostKondo} can be written as a Kondo Hamiltonian, we rewrite this operator in the form 
\begin{align}
H_\text{eff}=
\frac{1}{2}(S_+K_- + S_-K_+) + S_zK_z,
\end{align}
where $S_{\pm}$ and $S_z$ are the impurity spin operators, and  the $K$-operators act only on the bath. To obtain the operator $K$, we determine their matrix elements in the basis $|\sigma,n\rangle$, where $|n\rangle$ is some eigenstate of the free bath, $|\sigma\rangle$ is the state on the singly occupied impurity, and $|0,n\rangle$ denotes the states where the impurity is empty. For $K_-$ we get
\begin{align}
&\langle n |K_-|n'\rangle
=
\langle \uparrow,n |H_\text{eff}|\downarrow,n'\rangle
\nonumber\\
&
\hspace{5mm}
=
\frac{\alpha^2}{2}
\sum_{l}
\sum_{m}\Big[
\langle \uparrow,n|
\St^{(-l)}_+
|0,m\rangle  \langle 0,m|
V^{(l)}_-
|\downarrow,n'\rangle
\nonumber\\
&
\hspace{9mm}
-  
\langle \uparrow,n|
V^{(l)}_+ 
|0,m\rangle  \langle 0,m|
\St^{(-l)}_- 
|\downarrow,n'\rangle\Big],
\end{align}
and analogous expressions are obtained for $K_+$ and $K_z$. Using the explicit form \eqref{S1} for $\St_1$, we have 
\begin{align}
\langle n |
K_-|n'\rangle
&
=
\frac{\alpha^2}{2}
\sum_{lm}\Bigg[
\frac{
\langle \uparrow,n|
V^{(-l)}_+
|0,m\rangle  \langle 0,m|
V^{(l)}_-
|\downarrow,n'\rangle
}
{E_n-E_m-l\Omega}
\nonumber\\
&
-  
\frac{
\langle \uparrow,n|
V^{(l)}_+ 
|0,m\rangle  \langle 0,m|
V^{(-l)}_- 
|\downarrow,n'\rangle}
{E_m-E_{n'}-l\Omega}\Bigg].
\end{align}
A change of the summation variable $l\to-l$ in the second term gives 
\begin{align}
\frac{\alpha^2}{2}
\sum_{lm}
&
\langle \uparrow,n|
V^{(-l)}_+ 
|0,m\rangle  \langle 0,m|
V^{(l)}_- 
|\downarrow,n'\rangle
\,\,\times
\nonumber\\
&
\times\,\,\Big(
\frac{
1}
{E_n-E_m-l\Omega}
+
\frac{
1}
{E_{n'}-E_m-l\Omega}
\Big).
\end{align}

We now proceed separately for the odd end even terms in $l$, $K\equiv K_{o} + K_{e}$, where $K_{o,e}= \sum_{l=\text{odd},\text{even}} \cdots$. For both even and odd bath states [Eq.~\eqref{b-basis}] we expand $b_{0, e(o),\sigma} = \sum_k t_k b_{k,e(o) \sigma}$, where $k$ are eigenstates of the lead Hamiltonian. Using \eqref{hybodd} and \eqref{hybeven} we get
\begin{align}
&\langle n |K_{o,-}|n'\rangle =
\alpha^2v^2
\sum_{l\text{~odd}}
J_{l}(\eta)J_{-l}(\eta)
\nonumber\\
&\times
\sum_{m}
\sum_{kk'}
t_k
t_{k'}^*
\langle \uparrow,n|
c_{\uparrow}^\dagger
b_{k,o,\uparrow}
|0,m\rangle  \langle 0,m|
-b_{k',o,\downarrow}^\dagger
c_{\downarrow}
|\downarrow,n'\rangle
\nonumber\\
&\times 
\Big(
\frac{
1}
{E_n-E_m-l\Omega}
+
\frac{
1}
{E_{n'}-E_m-l\Omega}
\Big).
\end{align}
Using $J_{-l}(x)=-J_{l}(x)$ for odd $l$, and inserting the energy of the (only possible) intermediate state, we have
\begin{align}
&\langle n |K_{o,-}|n'\rangle =
\alpha^2v^2
\sum_{l\text{~odd}}
J_{|l|}(\eta)^2
\sum_{kk'}
t_k
t_{k'}^*
\nonumber\\
&
\times \langle n|
b_{k,o,\uparrow}
b_{k',o,\downarrow}^\dagger
|n'\rangle
\Big(
\frac{
1}
{\epsilon_f -\epsilon_{k'}-l\Omega}
+
\frac{
1}
{\epsilon_f -\epsilon_k-l\Omega}
\Big).
\end{align}
If we omit the matrix elements, the equation now expresses an operator identity.  We furthermore make the conventional assumption that mainly electrons close to the Fermi energy participate in the screening process, so that we can approximate  $\epsilon_{k},\epsilon_{k'}\approx E_f\equiv 0$ (energy zero) in the denominator. This leads to  (the minus sign is from permuting the operators)
\begin{align}
K_{o,-} =
-
2v^2
\sum_{l\text{~odd}}
J_{|l|}(\eta)^2
(s_{o,0})_-
\frac{
1}
{\epsilon_f -l\Omega},
\end{align}
where $(s_{o,0})_-=b_{o,0,\downarrow}^\dagger
b_{o,0,\uparrow}$ is the bath-spin operator on site $0$.
Because $\epsilon_f$ is negative, we have
\begin{align}
K_{odd,-} 
&=
\frac{J_{ex,o}}{2}(s_{o,0})_-,
\\
J_{ex,o}
&=
4v^2
\sum_{l\text{~odd}}
\frac{ J_{|l|}(\eta)^2}{|\epsilon_f| +l\Omega}.
\end{align}
An analogous calculation gives
\begin{align}
K_{even,-} 
&=
\frac{J_{ex,e}}{2}(s_{e,0})_-,
\\
J_{ex,e}
&=
4v^2
\sum_{l\text{~even}}
\frac{ J_{|l|}(\eta)^2}{|\epsilon_f| +l\Omega}.
\end{align}
Because of rotational symmetry in spin space, the other spin components must come out accordingly. In summary, the final Hamiltonian of the driven Anderson impurity model comes out as the asymmetric two-channel Kondo model, Eq.~\eqref{2chK} and \eqref{jbessel} of the main text.
We remark once again that the driving frequency should not allow for direct resonant excitations. 


\begin{thebibliography}{99}
 
\newcommand{\mytitle}[1]{``#1'',}

\bibitem{Goldman2014}
N.~Goldman and J.~Dalibard,
  \mytitle{Periodically Driven Quantum Systems: Effective Hamiltonians and Engineered Gauge Fields}
Phys. Rev. X {\bf 4}, 031027 (2014).


 \bibitem{Oka2009}
T.~Oka and H.~Aoki,
\mytitle{Photovoltaic Hall effect in graphene}
Phys. Rev. B {\bf 79}, 081406 (2009).

\bibitem{Kitagawa2011}
T.~Kitagawa, T.~Oka, A.~Brataas, L.~Fu, E.~Demler,
\mytitle{Transport properties of nonequilibrium systems under the application of light: Photoinduced quantum Hall insulators without Landau levels}
Phys. Rev. B {\bf 84}, 235108  (2011).

\bibitem{Jotzu2015}
G.~Jotzu, M.~Messer, R.~Desbuquois, M.~Lebrat, Th.~Uehlinger, D.~Greif, and T.~Esslinger, 
\mytitle{Experimental realization of the topological Haldane model with ultracold fermions}
Nature  {\bf 515}, 237 (2014).

\bibitem{Struck2011}
J.~Struck, C.~\"Olschl\"ager, R.~Le~Targat, P.~Soltan-Panahi, A.~Eckardt, M.~Lewenstein, P.~Windpassinger, K.~Sengstock
\mytitle{Quantum Simulation of Frustrated Classical Magnetism in Triangular Optical Lattices}
Science {\bf 333}, 996 (2011).

\bibitem{Else2016}
D.~V.~Else, B.~Bauer, and C.~Nayak,
\mytitle{Floquet Time Crystals}
Phys. Rev. Lett. {\bf 117}, 090402 (2016).

\bibitem {Wang2013}
Y.~H.~Wang, H.~Steinberg, P.~Jarillo-Herrero, and N.~Gedik,
\mytitle{Observation of Floquet-Bloch States on the Surface of a Topological Insulator}
Science {\bf 342}, 453 (2013).

\bibitem{Mentink2015}
J.~H.~Mentink, K.~Balzer, and M.~Eckstein,
\mytitle{Ultrafast and reversible control of the exchange interaction in Mott insulators}
Nature Comm. {\bf 6}, 6708 (2015).
	
\bibitem{Mikhaylovskiy2015}
R.~V.~Mikhaylovskiy, E.~Hendry, A.~Secchi, J.~H.~Mentink, M.~Eckstein, A.~Wu, R.~V.~Pisarev, V.~V.~Kruglyak, M.~I.~Katsnelson, Th.~Rasing, and A.~V.~Kimel, 
\mytitle{Ultrafast optical modification of exchange interactions in iron oxides}
Nature Comm. {\bf 6}, 8190 (2015).

\bibitem{Arenas2017}
J.~J.~Mendoza-Arenas, F.~J.~Gomez-Ruiz, M.~Eckstein, D.~Jaksch, S.~R.~Clark,
\mytitle{Ultra-fast control of magnetic relaxation in a periodically driven Hubbard model}
arXiv:1701.04123.

\bibitem{Eckstein2017}
M.~Eckstein, J.~H.~Mentink, Ph.~Werner,
\mytitle{Designing spin and orbital exchange Hamiltonians with ultrashort electric field transients}
arXiv:1703.03269.

\bibitem{Knap2016}
M.~Knap, M.~Babadi, G.~Refael, I.~Martin, E.~Demler,
  \mytitle{Dynamical Cooper pairing in nonequilibrium electron-phonon systems}
Phys. Rev. B {\bf 94}, 214504 (2016).

\bibitem{Komnik2016}
A.~Komnik and M.~Thorwart,
\mytitle{BCS theory of driven superconductivity}
Euro.~Phys.~Jour.~B {\bf 89}, 244 (2016).

\bibitem{Coulthard2016}
J. Coulthard, S. R. Clark, S. Al-Assam, A. Cavalleri, D. Jaksch,
\mytitle{Enhancement of super-exchange pairing in the periodically-driven Hubbard model}
arXiv:1608.03964.


\bibitem{Nozieres1980}
P.~Nozi\'eres and A.~Blandin, 
\mytitle{Kondo effect in real metals}
J.~de Physique {\bf 41}, 193 (1980).


\bibitem{Cox1998}
D.~L.~Cox and  A.~Zawadowski,
\mytitle{Exotic Kondo Effects in Metals: Magnetic Ions in a Crystalline Electric Field and Tunneling Centers}
Adv.~Phys. {\bf 47}, 599 (1998).

\bibitem{Oreg2003}
Y.~Oreg and D.~Goldhaber-Gordon, 
\mytitle{Two-channel Kondo effect in a modified single electron transistor}
Phys.~Rev.~Lett. {\bf 90}, 136602 (2003).

\bibitem{Potok2007}
R.~M.~Potok, I.~G.~Rau, H.~Shtrikman, Y.~Oreg and D.~Goldhaber-Gordon,
\mytitle{Observation of the two-channel Kondo effect}
Nature {\bf 446}, 167 (2007).

\bibitem{Keller2015}
A.~J.~Keller,	L.~Peeters,	C.~P.~Moca,	I.~Weymann,	D.~Mahalu,	V.~Umansky,	G.~Zar\'and, and  D.~Goldhaber-Gordon,
\mytitle{Universal Fermi liquid crossover and quantum criticality in a mesoscopic system}
Nature {\bf 526}, 237 (2015).

\bibitem{daSilva2009}
L.~G.~D.~da~Silva and E.~Dagotto,
\mytitle{Phonon-assisted tunneling and two-channel Kondo physics in molecular junctions}
Phys.~Rev.~B {\bf 79}, 155302 (2009).

\bibitem{Coleman2001}
P.~Coleman, C.~Hooley, and O.~Parcollet,
\mytitle{Is the Quantum Dot at Large Bias a Weak-Coupling Problem?}
Phys.~Rev.~Lett. {\bf 86}, 4088 (2001).

\bibitem{Rosch2001}
A.~Rosch, J.~Kroha, and  P.~W\"olfle,
\mytitle{The Kondo effect in quantum dots at high voltage: Universality and scaling}
Phys. Rev. Lett. {\bf 87}, 156802 (2001).

\bibitem{Krinner2015}
S.~Krinner,	D.~Stadler,	D.~Husmann,	J.-Ph.~Brantut, and 	T.~Esslinger,
\mytitle{Observation of quantized conductance in neutral matter}
Nature {\bf 517}, 64 (2015). 



\bibitem{Iwahori2016a}
K.~Iwahori and N.~Kawakami,
\mytitle{Long time asymptotic state of periodically driven open quantum systems}
Phys.~Rev.~B {\bf 94}, 184304 (2016).

\bibitem{Iwahori2016b}
K.~Iwahori and N.~Kawakami,
\mytitle{Periodically-driven Kondo impurity in nonequilibrium steady states}
Phys.~Rev.~A {\bf 94}, 063647 (2016).

\bibitem{Nakagawa2015}
M.~Nakagawa and  N.~Kawakami,
\mytitle{Laser-induced Kondo effect in ultracold alkaline-earth fermions}
Phys.~Rev.~Lett. {\bf 115}, 165303 (2015).


\bibitem{Bukov2015}
M.~Bukov, L.~D'Alessio, and A.~Polkovnikov,
\mytitle{Universal high-frequency behavior of periodically driven systems: from dynamical stabilization to Floquet engineering}
Adv. in Physics {\bf 64}, 139 (2015).

\bibitem{Eckardt2015}
A.~Eckardt and E.~Anisimovas,
\mytitle{High-frequency approximation for periodically driven quantum systems from a Floquet-space perspective}
New Jour. of Phys. {\bf 17}, 093039 (2015).
  
\bibitem{Mikami2016}
T.~Mikami, S.~Kitamura, K.~Yasuda, N.~Tsuji, T.~Oka, and H.~Aoki, Hideo,
\mytitle{Brillouin-Wigner theory for high-frequency expansion in periodically driven systems: Application to Floquet topological insulators}
Phys. Rev. B {\bf 93}, 144307 (2016).


\bibitem{Canovi2016}
E.~Canovi, M.~Kollar, and M.~Eckstein,
\mytitle{Stroboscopic prethermalization in weakly interacting periodically driven systems}
Phys. Rev. E {\bf 93}, 012130 (2016).

\bibitem{Bukov2015PRL}
M.~Bukov, S.~Gopalakrishnan, M.~Knap, and E.~Demler,
\mytitle{Prethermal Floquet Steady States and Instabilities in the Periodically Driven, Weakly Interacting Bose-Hubbard Model}
Phys.~Rev.~Lett. {\bf 115}, 205301 (2015).

\bibitem{Murakami2017}
Y.~Murakami, N.~Tsuji, M.~Eckstein, and P.~Werner,
\mytitle{Nonequilibrium steady states and transient dynamics of superconductors under phonon driving},
arXiv:1702.02942.

\bibitem{Babadi2017}
M.~Babadi, M.~Knap, I.~Martin, G.~Refael, and E.~Demler,
\mytitle{The theory of parametrically amplified electron-phonon superconductivity}
arXiv:1702.02531.


\bibitem{Sieberer2016}
L.~M.~Sieberer, M.~Buchhold, and S.~Diehl,
\mytitle{Keldysh Field Theory for Driven Open Quantum Systems}
Rep.~Prog.~Phys. {\bf 79}, 096001 (2016).

\bibitem{Sota2016}
S.~Kitamura and H.~Aoki,
  \mytitle{$\ensuremath{\eta}$-pairing superfluid in periodically-driven fermionic Hubbard model with strong attraction}
Phys. Rev. B {\bf 94}, 174503 (2016).

\bibitem{Bukov2016}
M.~Bukov, M.~Kolodrubetz, and A.~Polkovnikov,
\mytitle{Schrieffer-Wolff Transformation for Periodically Driven Systems: Strongly Correlated Systems with Artificial Gauge Fields}
Phys. Rev. Lett.  {\bf 116}, 125301 (2016).

\bibitem{aoki2014rev}
H.~Aoki, N.~Tsuji, M.~Eckstein, M.~Kollar, T.~Oka, and Ph.~Werner,
  \mytitle{Nonequilibrium dynamical mean-field theory and its applications}
Rev. Mod. Phys. {\bf 86}, 779 (2014).

\bibitem{eckstein2010}
M.~Eckstein and Ph.~Werner,
  \mytitle{Nonequilibrium dynamical mean-field calculations based on the noncrossing approximation and its generalizations}
Phys. Rev. B {\bf 82}, 115115 (2010).

\bibitem{Cox1993}
D. L. Cox and A. E. Ruckenstein, 
\mytitle{Spin-flavor separation and non-Fermi-liquid behavior in the multichannel Kondo problem: A large-N approach}
Phys. Rev. Lett. {\bf 71}, 1613 (1993).

\bibitem{Andrei1984}
N.~Andrei and C.~Destri 
Solution of the Multichannel Kondo Problem
Phys.~Rev.~Lett. {\bf 52}, 364 (1984).

\bibitem{Tsvelick1984}
A.~M.~Tsvelick and P.~B.~Wiegmann,
\mytitle{Solution of then-channel Kondo problem (scaling and integrability)}
Z.~Phys.~B {\bf 54}, 201 (1984).

\bibitem{Sacramento1991}
P.~D.~Sacramento and P.~Schlottmann,
\mytitle{Low-temperature properties of a two-level system interacting with conduction electrons: An application of the overcompensated multichannel Kondo model}
Phys. Rev. B {\bf 43}, 13294 (1991).


\bibitem{Strohmaier2010}
N.~Strohmaier, D.~Greif, R.~J\"ordens, L.~Tarruell, H.~Moritz, T.~Esslinger, R.~Sensarma, D.~Pekker, E.~Altman, and E.~Demler,
\mytitle{Observation of Elastic Doublon Decay in the Fermi-Hubbard Model}
Phys.~Rev.~Lett. {\bf 104}, 080401 (2010).

\end{thebibliography}
\end{document}